# Global Transients in ultraviolet and red-infrared ranges from data of the "Universitetsky-Tatiana-2" satellite.

G. K. Garipov[1], B.A. Khrenov[1], P.A. Klimov[1], V.V. Klimenko[2], E.A. Mareev[2], O. Martines[3], V.S. Morozenko[1], M.I. Panasyuk[1], I.H. Park[4], E. Ponce[3], H. Salazar[3], V.I. Tulupov[1], N.N. Vedenkin[1], I.V. Yashin[1].

[1] D.V. Skobeltsyn Institute of Nuclear Physics of M.V. Lomonosov Moscow State University, Leninskie Gory, Moscow,1, str. 2, 119991, Russia.

[2] Institute of Applied Physics of RAS, Ulyanov street, 46, Nizniy Novgorod, 603950, Russia.

[3] Department of Physics and Mathematics, University of Puebla, C.P. 72570, Puebla, Mexico.

[4] Research Center of MEMS Space Telescope of Ewha Womans University, 11-1 Daehyun-dong, Seodaemun-gu, Seoul, 120-750, Korea.

**Abstract**

Detectors of fast flashes (duration of 1-128 ms) in near ultraviolet (240-400 nm) and red-infrared (>610 nm) ranges on board the "Universitetsky-Tatiana-2" satellite have measured transient luminous events global distribution. Events with number of photons $10^{20}$-$5 \cdot 10^{21}$ radiated in the atmosphere are uniformly distributed over latitudes and longitudes. Events with number of photons more than $5 \cdot 10^{21}$ are concentrated near the equator above continents. Measured ratio of photons number radiated in red-IR range photons number radiated in UV related to excitation of nitrogen molecular indicates a high altitude (>50 km) of the atmospheric electric discharges responsible for the observed transients. Series of every minute transients (from 3 to 16 transients in the series) were observed. The detection of transients out of thunderstorm area, in cloudless region- sometimes thousands km away of thunderstorms is remarkable. The obtained data allow us to assume that transient events are not only consequences of lightning in event-by-event way but they are the result of "long distance" influence of thunderstorm electric activity causing breakdowns in the upper atmosphere (at altitudes >50 km).

## 1. Space detectors of the atmospheric ultraviolet and red-IR radiation.

Moscow State University satellite "Universitetsky-Tatiana-2" was launched on September 20 of 2009 to solar synchronous polar orbit with the height 820-850 km and inclination 98,8º [1]. The main goal of the scientific program of this satellite as well as of the first satellite of the same class "Universitetsky-Tatiana" [2] is the near Earth space study by measuring the charged particles fluxes at the orbit and radiation from the atmosphere. The data discussed in this paper were collected in period from October 20, 2009 to January 17, 2010.

Radiation from the atmosphere was measured by detectors in two ranges of wavelength: 240-400 nm (near ultraviolet, UV) and >610 nm (red-infrared, RI). Both detectors are the photomultiplier tubes of Hamamatsu type R1463 with multialcali cathode and borosilicat window covered by different filters: UVS1 (240-400 nm) and KS11(>610 nm). It is worthwhile to note that quantum efficiency of PM tube cathode is uniform in UV range (QE~20%) and it is rapidly decreasing in RI range (QE=6% at 600 nm and QE=1% at 800 nm). Field of view of detectors is oriented to nadir and determined by collimator (holes of diameter 0.8 mm in the black 2.2 mm thick cover). From the orbit height 850 km those detectors observe the atmosphere area with effective diameter 300 km (effective area $7·10^4$ km$^2$).

The conversion of analog signals to digits in both detectors was done every microsecond by one ADC with the help of the multiplexor. Much larger waveform sample (1 ms) was applied for recording the temporal 128 ms event profile by summing digital signals measured every microsecond. Data on 128 ms trace were recorded in operative memory and every millisecond the new trace data were compared with the previous one. If the maximal 1 ms signal in the new trace was larger than in the previous one the new trace was recorded as the reference trace.

Every minute the trace with the largest 1 ms signal was rerecorded in the main memory. The main memory data was transmitted to the mission center when satellite flied over Moscow region.

The gain of both PM tubes was controlled by the atmospheric UV glow so that the average ADC signal in UV was kept constant (ADC code N=16) with the relaxation time of 0.25 sec. The PM tubes gain was recorded as code M of digital- analog convertor controlling the PMT voltage. During the time interval of 128 ms transient event duration the gain was constant. For more detailed description of the detectors electronics see [1, 2, 3].

## 2. Transient events characteristics.

The first results of transient events observation by detectors of the "Unversitetsky-Tatiana-2" satellite were published elsewhere [1,4]. In this paper we present more detailed data with larger statistics.

Every minute UV detector selects the event with maximal 1 ms signal in analysis of 468 measured traces. If in this period there is no transient atmospheric phenomenon the UV detector is triggered by maximal signal expected from statistical fluctuations of the average photon flux received in the detector field of view. Statistically every minute a signal of about $5\sigma$ expected as the largest one ($\sigma$ is the standard deviation from the average). Signals much larger than $5\sigma$ are rare (for example, one signal more than $10\sigma$ is expected in $10^5$ minutes). In the experiment, the signal more than $10\sigma$ occurs every 10 minutes in average, and every 1 minute- over some regions of the Earth. It shows that the triggering system selects mostly the real transient events but not the statistical fluctuations of atmosphere glow.

Calibration of the detector signal in number of photo electrons in PM tube cathode was done before flight in measurements of single photo electrons [3]. Hamamatsu R1463 tube selected for the detector has a good resolution in single photo electrons. For conversion from number of photo electrons really measured by PM tube to photons coming from the atmosphere the values of quantum efficiency presented by Hamamatsu [5] was used (20% for UV detector and average value of 2% - for RI detector).

Every transient event was presented by number of photons Q, measured during 128 msec. In assumption of isotropic fluorescence radiation of the event source number of photons $Q_a$ radiated in the atmosphere was calculated as

$$Q_a = Q \, 4\pi R^2/S, \tag{1}$$

where S – is operation area of the PM tube cathode, R- is the distance between the atmosphere and the detector. For UV detector S=0,4 cm$^2$, for RI detector the cathode area was restricted to S=0,2 cm$^2$. For R~800 km the ratio $Q_a/Q=2.2\cdot10^{17}$ for UV detector.

For comparison with other results on atmospheric transients the energy released in UV or RI photons in the atmosphere was calculated as

$$E = \varepsilon \times Q_a, \tag{2}$$

where $\varepsilon$ – is the average photon energy (for UV range $\varepsilon$=3.5 eV, for RI range $\varepsilon$=1.75 eV).

In period October 2009 –January 2010 the «Universitetsky-Tatiana-2» detectors orbited the Earth 797 times with 320 hours of operation time in shadowed "night" part. In off- line analysis of the experimental data 2628 transient events were selected. In this analysis additional selection criteria were applied: in ADC codes the maximal UV signal measured in 1 ms time sample has to be N>80.

Selected transient events found to be distributed in wide range of photons numbers: $Q_a \sim 10^{20}$-$10^{26}$ (energies in UV $E_{uv} \sim 5.6 \cdot 10^1$-$5.6 \cdot 10^7$ J). The observed range of photons number is much wider than the available range of ADC codes in one time sample (1-1024) due to: a) summing up data from all 128 time samples numbers in one trace, and b) measurement of signals at variable PMT's gain (variable intensity of atmosphere glow).

Transients differential and integral distribution on photons number $Q_a$ are presented in Fig.1 and 2. Maximum in differential $Q_a$ distribution indicates the experimental event threshold: $Q_{a(thr)} \sim$

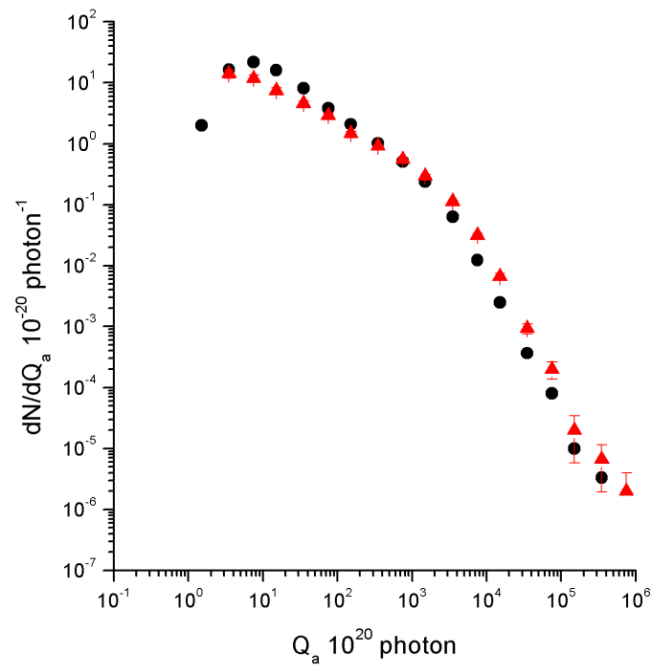

Fig. 1 Differential distribution of transients on number of photons $Q_a$. Circles are numbers of photons in UV range. Triangles –numbers of photons in RI range.

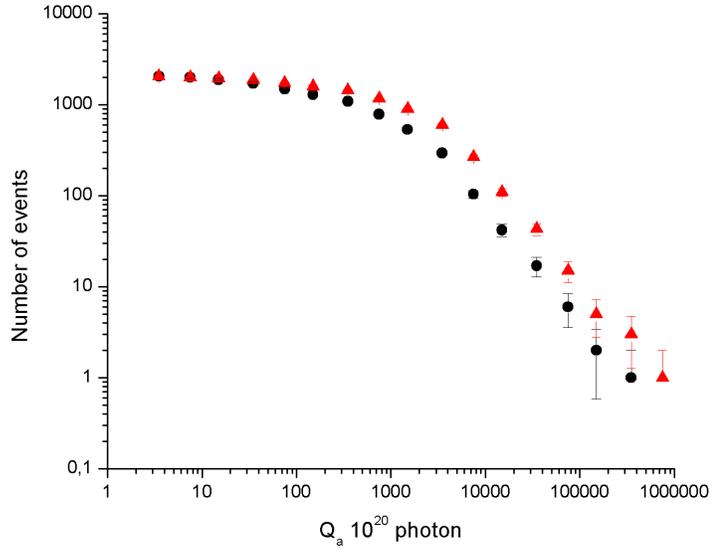

Fig. 2 Integral distribution of transients on number of photons $Q_a$. Circles and triangles as in Fig.1.

~3 $10^{20}$ ($E_{uv}$~170J). In range of $10^{21}<Q_a<10^{23}$ differential distribution could be approximated by power law with exponent $\gamma = -1$. Starting from $Q_a = 10^{23}$ the power law exponent changes to $\gamma = -2$. Temporal profiles of transient events vary from single millisecond (ms) flashes to longer (up to 128 ms) flashes with variable structure. Profile examples are presented in Fig. 3:

a – short single flash, b – repeating short flashes, c- structural longer flash. There is a general tendency in correlation of profile structure with photon number in a transient: events with $Q_a < 3 \cdot 10^{21}$ are mostly short (a-type in Fig.3), events with the large photon numbers $Q_a > 10^{24}$ are mostly long with duration of tens- to- hundred ms (c- type in Fig. 3). In many intermediate events short pulses are repeated several times in trace of 128 ms (Fig. 3 b).

An interesting parameter of the transient event is the ratio P of photons number in RI range to photons number in UV range. P-ratio was measured for every 1 ms interval of the transient event and summed up to trace period 128 ms. A difficult point in such measurement is the saturation of ADC in measurement of UV signal profile (RI signals measured by ADC are ten times lower than in UV range due to lower quantum efficiency of PMT's cathode). To get the true P- ratio only UV events with ADC codes N in limited range were considered: 80<N<1024.

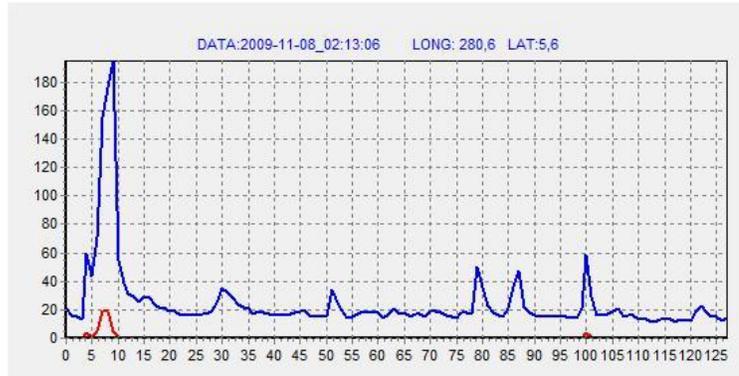

**A**

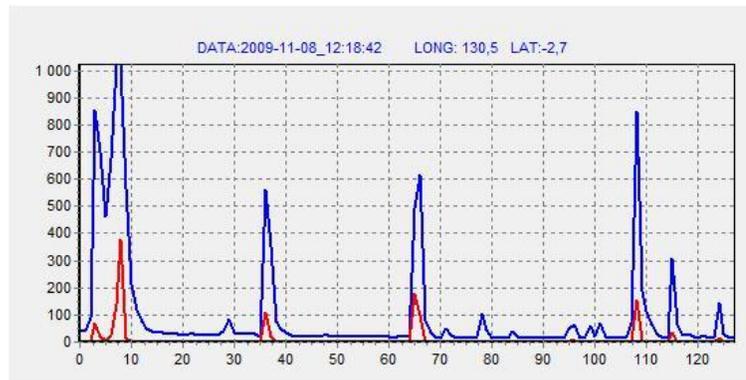

**B**

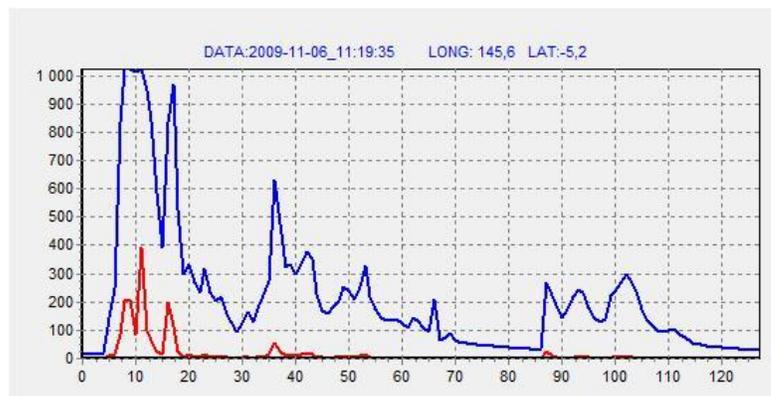

**C**

Fig. 3 Examples of ms transient temporal profiles. A- Short (1-5 ms) single pulse (secondary pulses add less than 50% to the integral photon number). B- Multiple short pulses. C- Longer profile.

For pulses with duration <5 ms the P-ratio distribution is presented in Fig. 4 by clear histogram. This distribution has maximum at P=3.6 with half maximum width ΔP=±1.5. P-ratio for photons summed over 128 ms trace is presented in Fig. 4 by dashed histogram. This distribution has maximum at P=1.5 with half maximum width ΔP=±1. Statistics of short pulses are larger than statistics of 128 ms traces due to events with several short pulses in one trace.

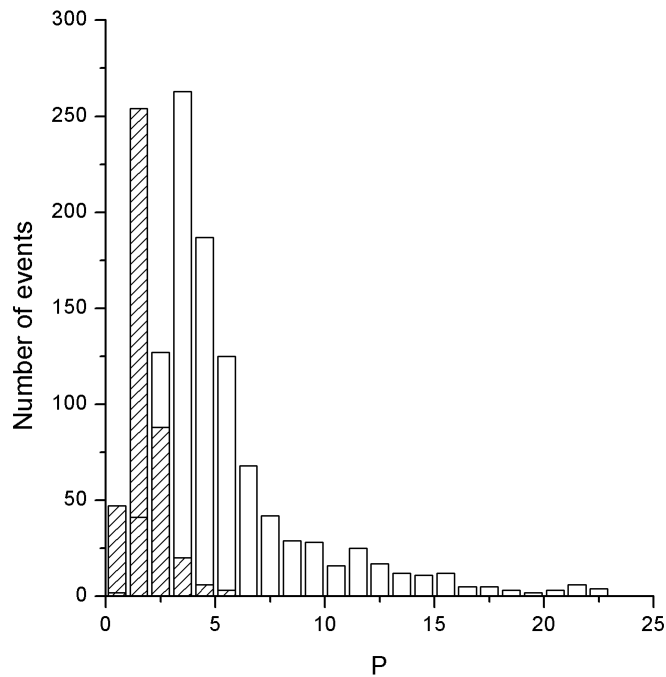

Fig. 4  P- ratio distribution. Open histogram- for photons in short (<5 ms) pulses. Dashed histogram- for photons summed over 128 ms trace.

The P-ratio as a measure of flash radiating spectrum gives some information on its origin. It is well known that lines of molecular nitrogen ($1PN_2$ and $2PN_2$ transitions) dominate in radiation of the atmosphere molecules and atoms excited by moderate (not lightning) electric discharges. In lightning (particularly in the brightest return stroke) atomic oxygen lines dominate. In Fig. 5 (adapted from [6]) an example of lightning return stroke spectrum is presented.  The P-ratio in this lightning radiation is of about 10 and it may be higher when radiation is measured from the satellite orbit, after light scattering and absorption in clouds. Experimentally measured P-ratio in flashes does not confirm the lightning spectrum. It is close to the expected ratio in moderate (streamer for example) electric discharges where radiation from molecular nitrogen prevails.

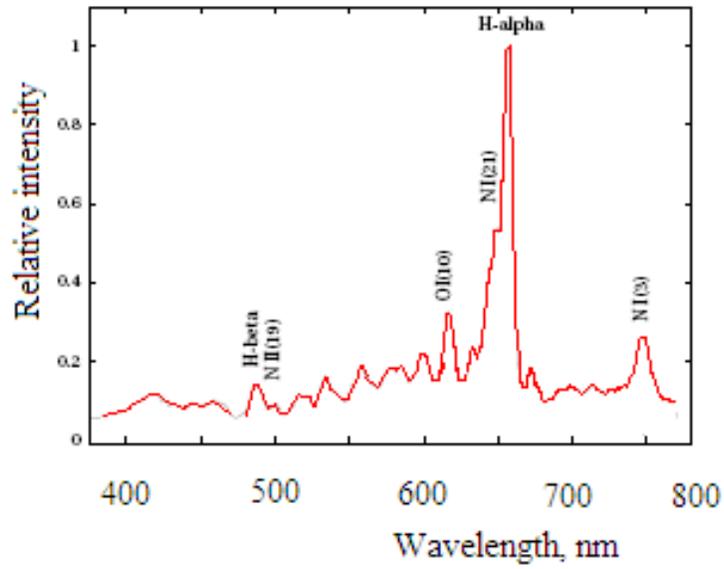

Fig. 5 Spectrum of lightning radiation (return stroke).

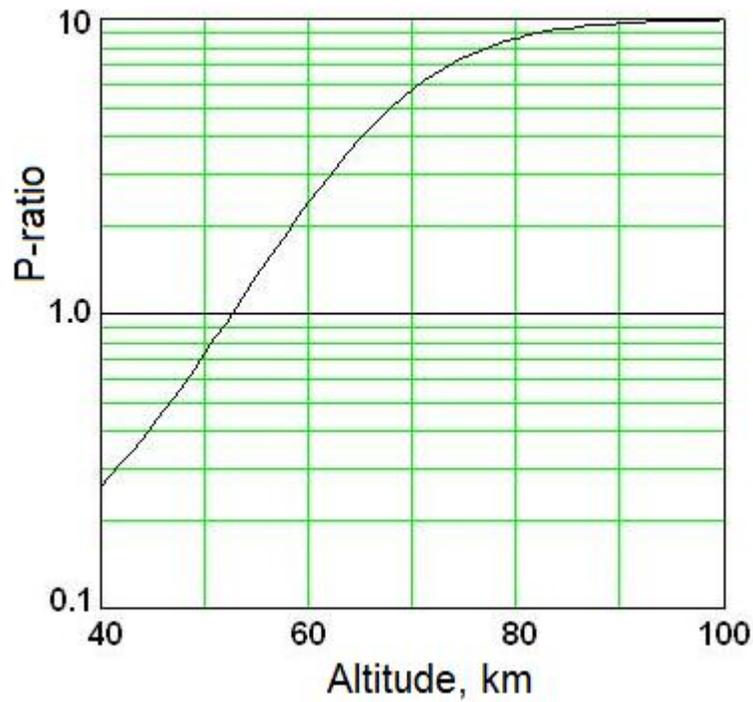

Fig. 6 P-ratio as function of altitude in the atmosphere (calculation).

Expected P-ratio in the atmospheric electric discharges was calculated for sum of 1PN$_2$ and 2PN$_2$ excitation states by the following formula:

$$P = \frac{Q_{aIR}}{Q_{aUV}} \approx \frac{q^*_{1PN_2}}{q^*_{2PN_2}} \cdot \frac{1+(\nu_d \tau_r)_{2PN_2}}{1+(\nu_d \tau_r)_{1PN_2}} \qquad (4)$$

where $q^*$ is the rate of corresponding transitions depending on excitation cross-sections and electron energy distribution. The characteristic value of $q^*_{1PN2} / q^*_{2PN2}$ ratio is of about 10. Radiation life time of excitation levels $1PN_2$ and $2PN_2$ are $\tau_r(1PN_2) \approx 8 \cdot 10^{-6}$ s and $\tau_r(2PN_2) \approx 9 \cdot 10^{-8}$ s. $\nu_d$ is the rate of collisions in which the excitation energy is lost without radiation, $\nu_d = \sigma \cdot V_T \cdot n_m(H)$ ($\sigma$ is collision cross-section assumed as kinetic $\sim 10^{-15} cm^2$), $V_T(H) = (4/3) \cdot [8kT(H)/\pi m]^{1/2}$ is the molecular velocity relative to average temperature velocity, $H$ is altitude in the atmosphere, m is nitrogen molecular mass. Molecular density was assumed as $n_m(H) = 1.8 \cdot 10^{15} \cdot exp[-(H-70)/H_o)]$ where $H_o = 7$ km is parameter of the exponential atmosphere. It was also assumed that atmosphere temperature $T(H)$ linearly decreases with altitude from 237 K at $H=50$ km to 173 K - at H= 80 km.

Calculated P-ratio as a function of the atmosphere altitude is presented in Fig. 6. Comparison of the experimental P-ratio distribution presented in Fig.4 with the calculated ratios indicates the range of altitudes $H$ in the atmosphere corresponding to the experimental P values: $H= 50-80$ km. The highest experimental P-ratios P>10 may be related to lightning.

### 3. Transient events global distribution.

Position of every transient event is determined in geographical coordinates by Universal Time (UT) of the event. In 642 of all 797 available "working" orbits there is at least one transient event. Global distribution of all transient events is presented in Fig. 7. Local night period of observation has distinct borders: latitude 60°N in the Northern and latitude 30°S - in the Southern Hemisphere. The distribution has an evident component concentrated above continents in equatorial regions of America, Africa and Indochina. Distribution of "zero"- orbits, presented in Fig. 8, demonstrates the same tendency: zero- orbits "avoid" the regions where the transient rate is high.

Trying to separate two components of transient event distribution: 1) uniform over globe and; 2) concentrated above equatorial regions, comparison of global distributions of events with various photon number $Q_a$ was made. The results of this analysis are presented in Fig. 9a and 9b. Events with photon numbers $Q_a < 5 \cdot 10^{21}$ shows less concentration to equator. Global distribution of events with photons numbers $Q_a > 5 \cdot 10^{21}$ demonstrates strong concentration at latitudes 0-30° N and S. In both components there is an interesting absence of transients: above Sahara and Australia deserts, and above Siberia.

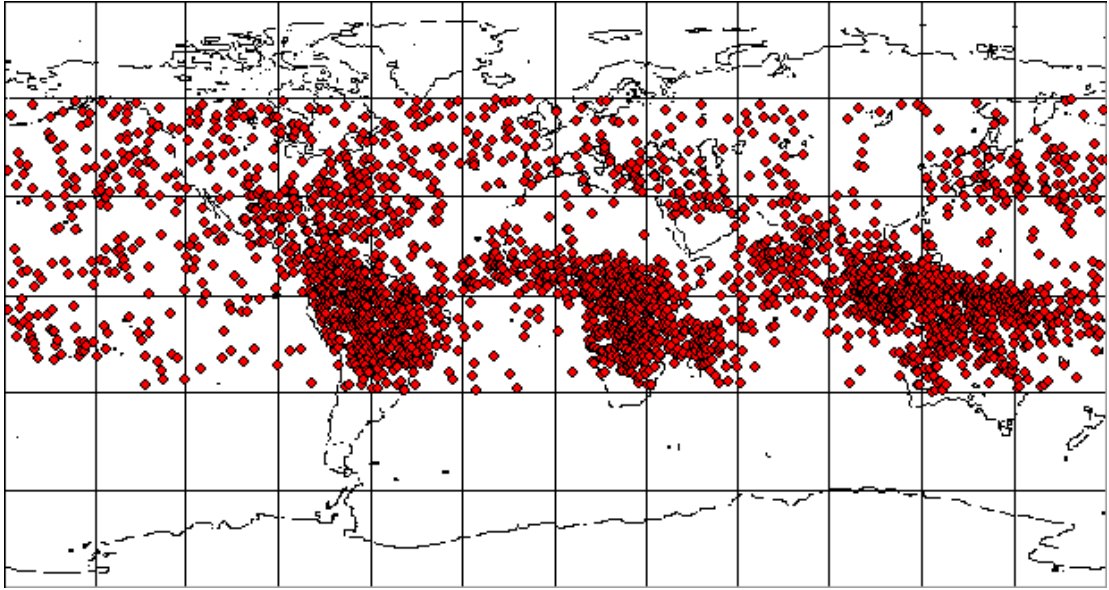

Fig. 7 Global distribution of all transient events. Points are coordinates of transient events.

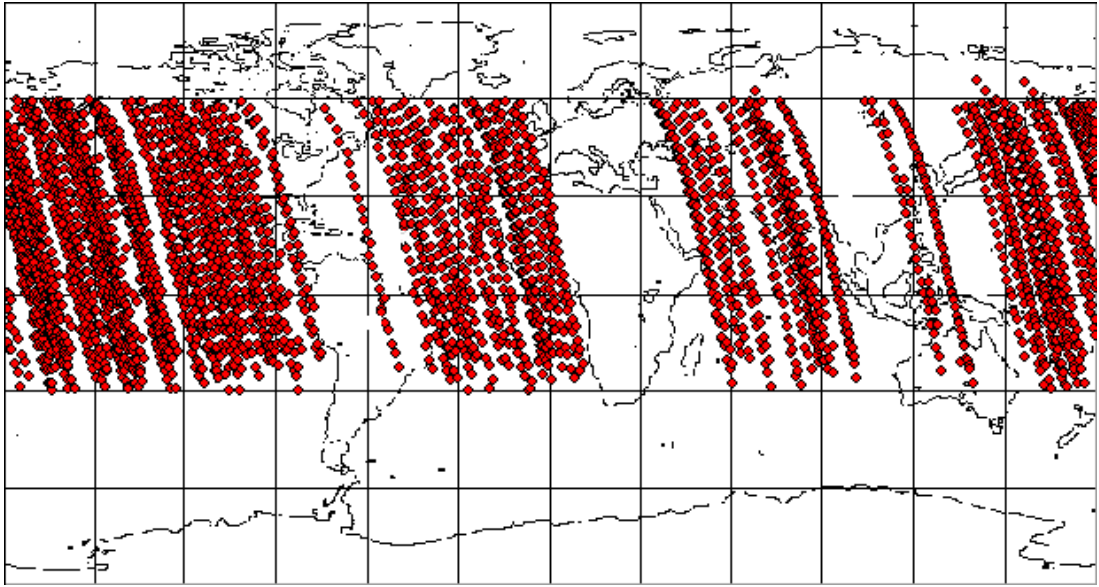

Fig. 8 Distribution of orbits without transients. Points are every minute coordinates of the satellite.

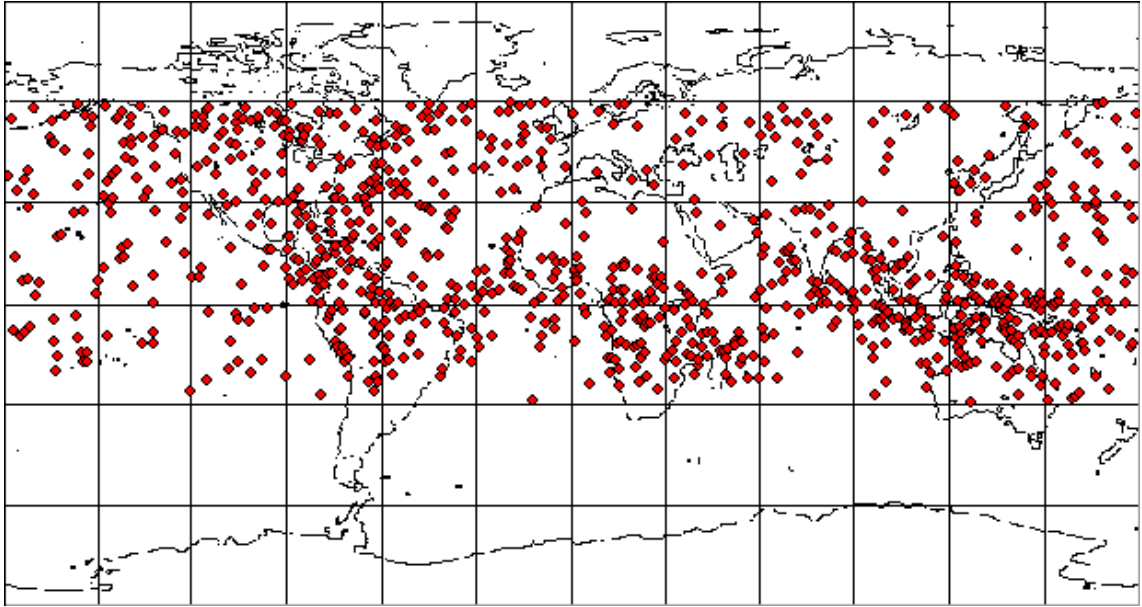

Fig. 9a Global map of UV transients with $Q_a < 5 \cdot 10^{21}$.

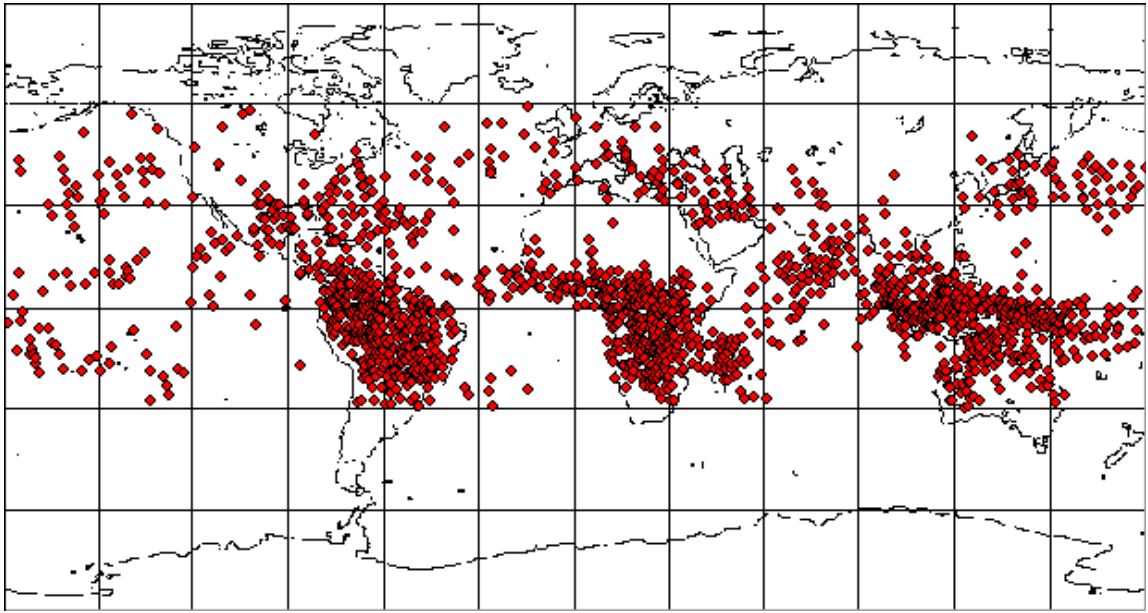

Fig. 9b. Global map of transients with $Q_a > 5 \cdot 10^{21}$.

It was found that many transients were detected above cloudless regions. For selection of clouds and cloudless regions the data from [7] were used where IR cloud map is available every 3 hours. Observing borders of large cloud region do not change much in 3 hours the "cloud" and "cloudless" regions were marked and a number of transient events in cloud and cloudless areas were compared. Among analyzed 1354 events number of transients in cloudless regions (627) was found to be near to number of transients in cloud regions (727). It is well known that lightning events are strongly correlated to cloud regions and it was confirmed by using data of [7] on clouds for comparison with lightning global distribution [8]. A large portion of transients observed in the present experiment in cloudless areas is difficult for understanding in assumption of event-to-event relation between transients and lightning. At the same time our data are not the first in observing the transients out of thunderstorm regions, see [9,10,11].

### 3. New phenomenon of series of transients.

In the "Universitetsky-Tatiana-2" satellite experiment a new feature of transient phenomenon was observed: sequences of transients registered every minute in one orbit, Fig. 10. A sequence of detected every minute $N_s \geq 3$ transients was called "series" as such sequences are statistically improbable for transients detected with the rate ~0.1 per minute in average. More than 50% (1519) of all transients (2628) were detected in series which shows the importance of this feature for the transients detected in the present experiment. Analysis of "single" and "serial" transient global distribution showed a large difference: single transient distribution is uniform (Fig. 11) while serial transients are concentrated in equatorial regions above continents (Fig. 12). Most of the single transients (90%) are short pulses (less than 10 ms) of type "a" and "b" presented above in Fig. 3. In serial transients percentage of short pulses is much less: 60%. There is also a difference between single and serial transients in their distribution over photon number $Q_a$, Fig. 13.

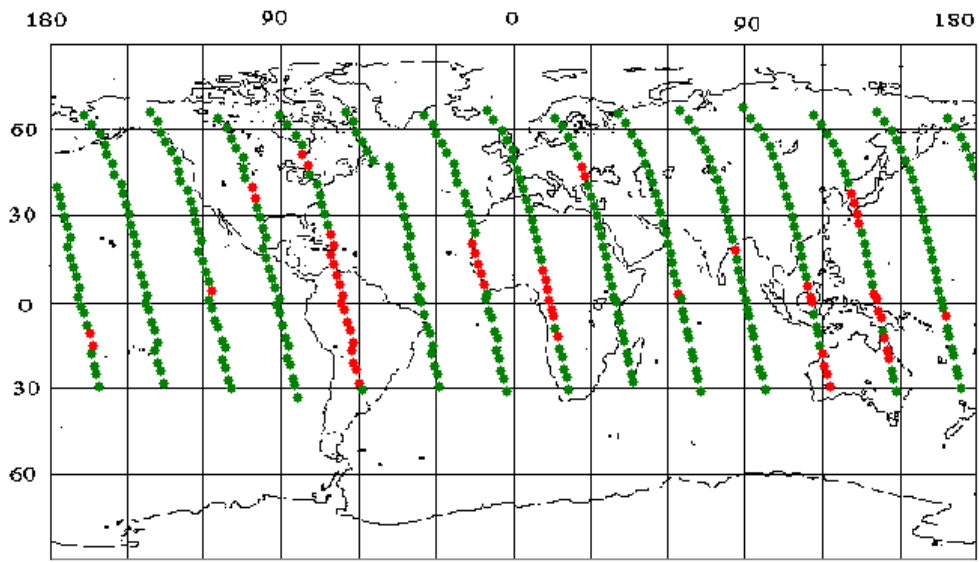

.

Fig.10. One day (14 orbits) of transients observation. Green points are every minute coordinates of the satellite (at local night in the latitude range from 30°S to 60°N) and red points are coordinates of detected transients.

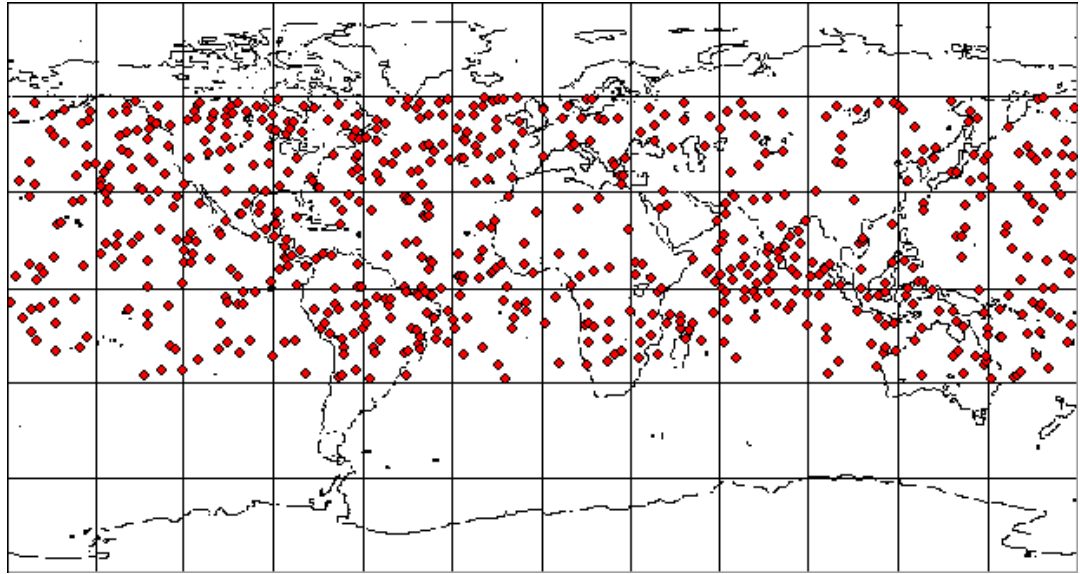

Fig. 11. Global distribution of "single" transients.

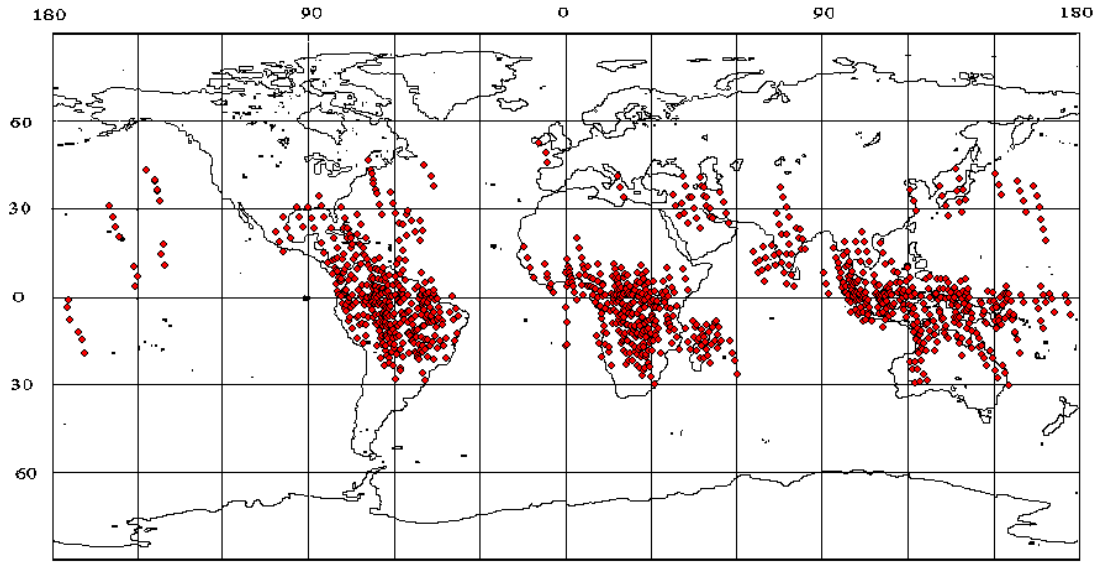

Fig. 12 Global distribution of "serial" transients.

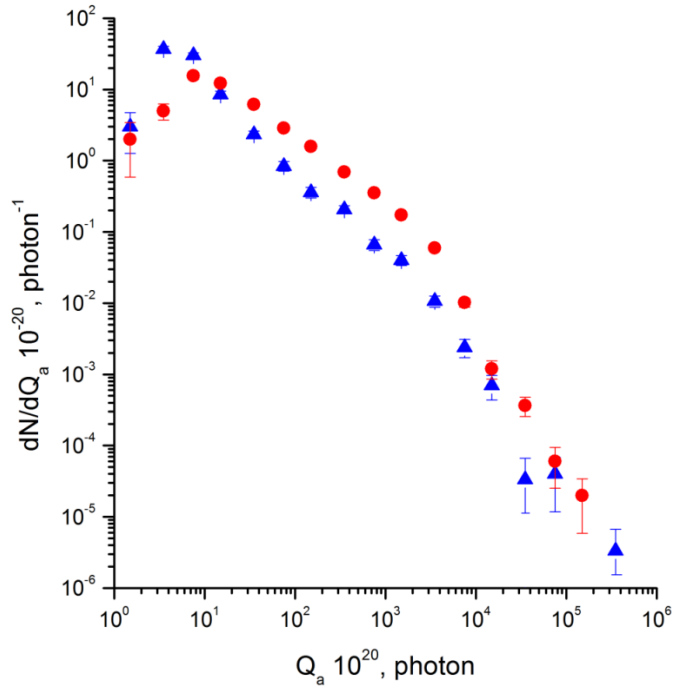

Fig.13 Differencial distribution of single (triangles) and serial (circles) transients over photon number in the atmosphere.

Among single transients there are less "bright" ones than among serial transients. This feature correlates with mentioned above (section 2) relation between transient duration and photon number: short flashes are generally less bright than longer ones. Transient distribution over number of transient in series $N_s$ and photon number $Q_{amax}$ of the brightest transient is presented in Table 1. Number of events in the first column is for single transient number ($N_s =1$). Single transients contain only 40% of events with $Q_a>5\cdot10^{21}$ while in series ($N_s \geq 3$) 75 % of events contain maximal transient photon number $Q_{amax} > 5\cdot10^{21}$. The higher is photon number $Q_{amax}$ the higher is probability to find long sequence of transients: for $Q_{amax} <5\cdot10^{21}$ probability to detect long sequence $N_s>4$ is 16% ; for $Q_{amax}>5\cdot10^{21}$ it is 25%.

Table 1. Distribution on number $N_s$ of transients in series, for two event groups with different $Q_{amax}$.

| $Q_{amax}$ | $N_s$ | | | | | | | | | | | | | | | |
|---|---|---|---|---|---|---|---|---|---|---|---|---|---|---|---|---|
| | 1 | 2 | 3 | 4 | 5 | 6 | 7 | 8 | 9 | 10 | 11 | 12 | 13 | 14 | 15 | Sum |
| $<5\cdot10^{21}$ | 321 | 62 | 34 | 29 | 13 | 11 | 14 | 8 | 1 | 2 | 1 | 2 | 2 | 0 | 1 | 501 |
| $>5\cdot10^{21}$ | 354 | 372 | 422 | 311 | 156 | 97 | 119 | 88 | 44 | 38 | 54 | 34 | 24 | 0 | 14 | 2127 |

One of the most interesting features of serial transients is that a large portion of them occurs in cloudless regions. In Fig. 14 a typical example of transient series in the region with a known cloud cover and lightning distribution is presented. The satellite goes from South to North and the first three transients were detected in the thunderstorm region. The next 3 transients are in cloudless region where lightning were not detected. Such observation contradicts the event-to-event relation between lightning and transient in this case. More likely transients in the atmosphere at altitudes 50-80 km (low ionosphere region) are atmospheric phenomenon induced by lightning from large distance. In the presented example, distance between lightning active zone and detected transients in a cloudless region is of about $10^3$ km. Transients in series with greater distances ($\sim 5 \cdot 10^3$ km) from thunderstorm regions were also detected (see an example of series in South America, Fig. 10).

The other transient series feature − transients follow geomagnetic lines- might be experimental artifact as the satellite orbits are close to geomagnetic meridians and direction across the meridians are investigated only at the next orbit which crosses the same latitude much later (90 minutes) and at distance $\sim 3 \cdot 10^3$ km away of the previous orbit. On the other hand, discovery of electron-positron fluxes trapped in the geomagnetic field [12] in correlation with gamma- transients indicates possible active role of geomagnetic field in making transient series.

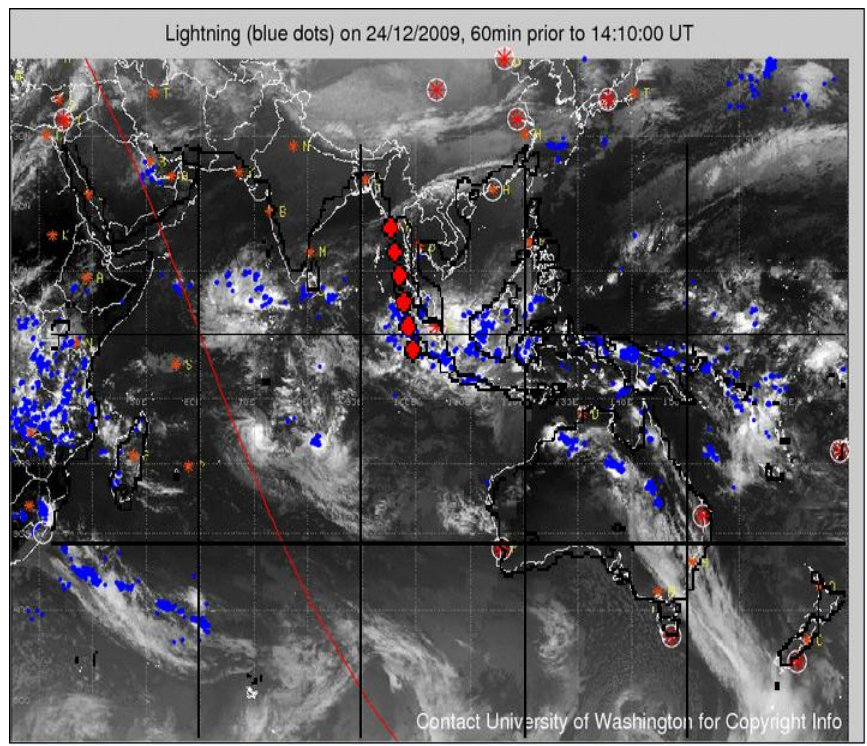

Fig. 14 Example of transient series detected in the region with a known cloud cover and lightning activity.

The rate of transients in series in various Earth areas were compared with the rate of lightning detected in LIS experiment [13]. In Fig. 15 (upper panel) the global map of the transient rate is presented (in number of events per square km per hour) for units in the atmosphere 10°×10° (note that our data is available only at local nights, i.e at latitudes from 60°N to 30°S). In the most active areas (South America, Africa, Indonesia-Australia) the transient rate in series is of about $3 \cdot 10^{-5}$ events/(km$^2$ hour). There are areas where series were not detected at all and some areas over oceans with rate $\sim 3 \cdot 10^{-6}$ events/(km$^2$ hour). The rate of transients in series and the rate of transients with photon number $Q_a > 5 \cdot 10^{21}$ (Fig. 9b) are close and their global distributions are similar (Fig. 9b and Fig. 12). Comparison with data of LIS experiment [13] on rate of lightning (Fig. 15, lower panel) shows that in the most active thunderstorm regions above continents the ratio of transient rate to lightning rate is T/L~0.5%. Interesting fact is that there are areas above oceans (North Atlantic, North Pacific, Indian oceans) where the ratio T/L~5% and both lightning and transient rates are higher than average for ocean regions.

The transient rate numbers in the present work are in agreement with ISUAL data [14] if all kind of TLE detected in ISUAL experiment are summed up. The difference of T/L ratios above continents and above specific area of oceans were also noticed in [14].

### 4. Discussion and Conclusion.

The "Universitetsky-Tatiana-2" detectors of atmospheric transients in UV and RI ranges allowed us to select and measure events with duration of 1-128 ms all over the globe in wide range of photon number $Q_a$, radiated in the atmosphere. Important feature of the experiment is observation in nadir direction and triggering by flashes in near UV (wavelengths 240-400 nm). Such triggering suppressed the rate of lightning occurring in the detector FOV. Observed P- ratio of Red-Infrared photon number to UV photon number in the range of 1<P<10 indicated a high altitude (50-100 km) of observed transients in the atmosphere.

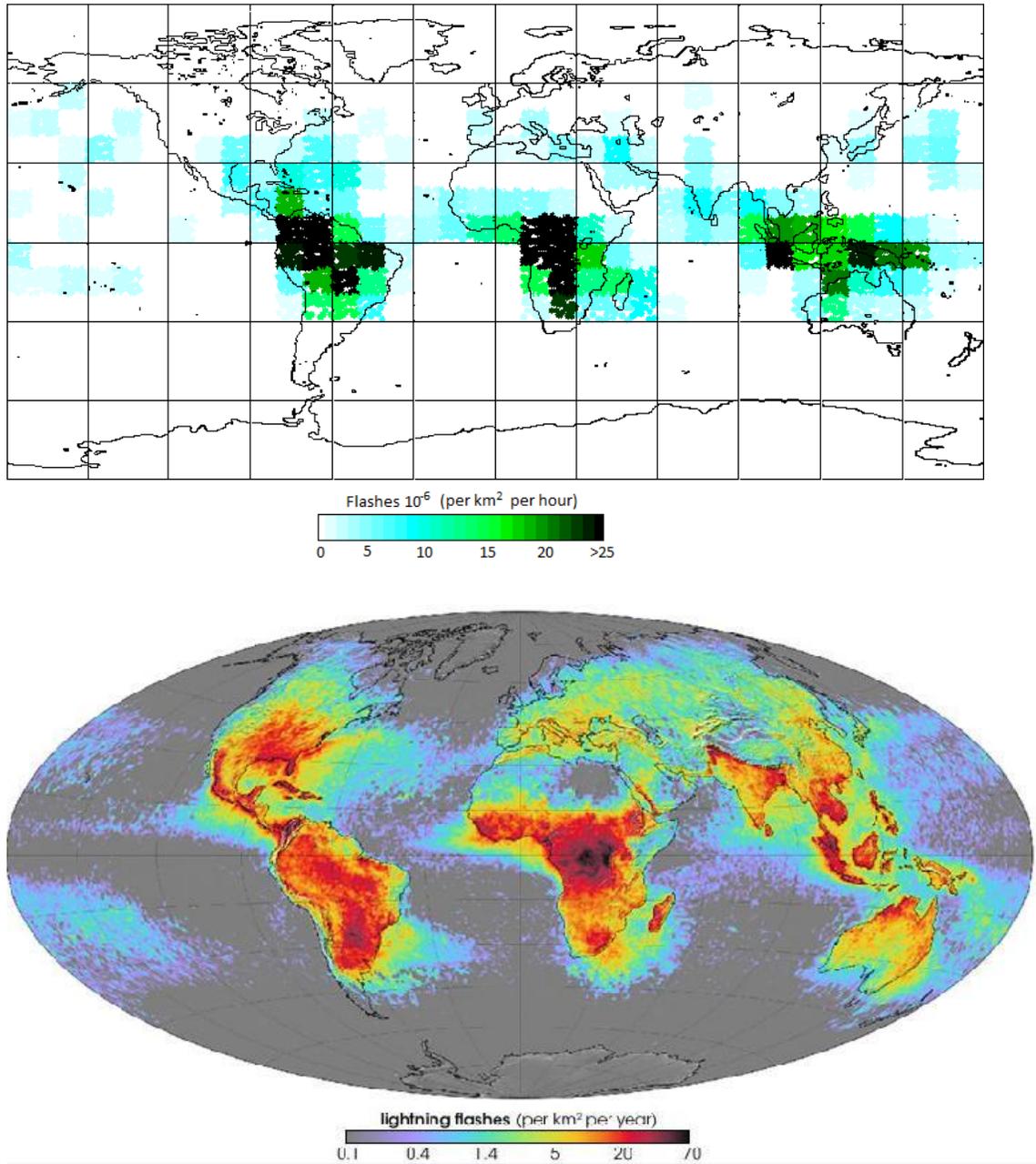

Fig. 15 Upper panel: global map of transient rate (Universitetsky-Tatiana-2 data). Bottom panel, global lightning rate.

Measured distribution of transients over photon numbers $Q_a$ shows two ranges with a changing exponent of power law approximating the differential distribution: $\gamma = -1$ for $10^{21} < Q_a < 10^{23}$ and $\gamma = -2$ for $Q_a > 10^{23}$. This change of exponents is observed at photon numbers far away of the detector threshold and indicates a physical threshold in production of transients (at UV photon number $Q_{athr} = 10^{23}$). In integral distribution this threshold is evident (Fig. 2). In assumption that transient photon number is in proportion to electric atmospheric discharge energy it means that discharges responsible for transients of $Q_a > 10^{23}$ are also responsible for released energy in the atmosphere.

Several kinds of correlations underline a difference between transients with "large" and "small" photon numbers: small transients are shorter in time, small and short transients are more uniformly distributed in the global map, large transients are concentrated near equator and above continents, large transients have a tendency to be observed in series, occurring every minute on satellite route.

The rate of observed transients is much less than the rate of lightning systematically detected in various Earth zones. The ratio of transient to lightning rate is ~0.5% in thunderstorm areas near the equator above continents. It is order of magnitude higher above oceans although the absolute rate of both transient and lightning events above oceans is much less than the rate above continents.

"Small" transients in photon numbers ($Q_a < 5 \cdot 10^{21}$) and "single" (not accompanied by transient in series) appeared as "independent" of lightning as their global distribution does not show any correlation with thunderstorm areas. Transients observed in cloudless areas far away of thunderstorms also could not be considered as correlated to lightning. Percentage of transients observed in cloudless areas is high (~50%) which shows a regular character of transient production outside the thunderstorms.

The presented experimental data on transients lead to conclusion that thunderstorms with lightning are responsible not only for lightning themselves but for general excitation of the upper atmosphere (50-100 km) at thousands km away of thunderstorm. Excitation of the atmosphere far away from the electromagnetic pulse transmitted from the powerful radio stations was long ago observed in experiments [15]. It is well known that the lightning electromagnetic pulse transfers for long distances and may cause "secondary" discharges [16]. Assuming that the electric field in a wave falls down linearly with distance to lightning and the photon number in "secondary" transient is in proportion to the arrived electric pulse, the distribution on photon numbers with exponent $\gamma = -1$ (as observed in the present experiment) may be obtained for a given "primary" lightning energy (see for details [1]). Exponent $\gamma = -2$ for larger photon numbers $Q_a > 10^{23}$ is an exponent of the electric pulse distribution of lightning responsible for "original" transients. In this scenario a threshold photon number $Q_a \sim 10^{23}$ divides "secondary" (small photon numbers) and "original" (large photon numbers) transients with their observed properties.

**Acknowledgement.**

Authors are grateful to A.V. Gurevich for helpful discussions. Support from RFFI grants 07-02-92004-ННС, 09-02-12162-ofi_m, 10-05-01045a are gratefully acknowledged.

**References.**

1. Sadovnichy V.A., M.I. Panasyuk, I.V. Yashin et al**.** Solar System Research, 45, №1, 3-29 (2011).
2. Sadovnichy V.A., M.I. Panasyuk, S.Yu. Bobrovnikov et al. Cosmic Research, 45, #4, 273-286 (2007).
3. Garipov G.K., M.I. Panasyuk, I.A. Rubinshtein, et al. Instruments and Experimental Techniques, 49, #1, 126-131 (2006).
4. Vedenkin V.V., G.K. Garipov, V.V. Klimenko, et al . JETP, 140, 900-910 (2011).
5. Hamamatsu, data book for PM tubes (1998).
6. Heavner M.J. University of Alaska, Fairbanks. PhD Thesis (2000).
7. Space Science and Engineering Center Images, NASA,www.ssec.wisc.edu/data/comp/ir
8. University of Washington, http://webflash.ess.washington.edu
9. Chen A.B. et al.  JGR, 113, A08306, doi:10.1029/2008JA013101 (2008).
10. Su H.T., Hsu R.R, Chen A.B., et al., Nature, 423, 974- 976, (2003).
11. Mende S.B. et al. In "Sprites, Elves and Intense Lightning Discharges", Ed.: M. Fullekrug,
12. Briggs, M. S., et al.  Geophys. Res. Lett., 38, L02808, doi:10.1029/2010GL046259, (2011).
13. Christian, H.J., et al., JGR , 108 (D1), 4005, doi:10.1029/2002JD002347 (2003).
14. Hsu R.R., A.B. Chen, C.L. Kuo et al,  in "Coupling of Thunderstorms and Lightning Discharges to Near Earth". Ed.: N.B. Crosby, N-Y. Huang and M.J. Rycroft. AIP1118, p. 99 (2009).
15. Gurevich A.V. Uspekhi Fizicheskikh Nauk, 177, №11, 1145-1177 (2007). doi:10.3367/UFNr.0177.200711a.1145.
16. Inan, U. S., S. A. Cummer, and R. A. Marshall.  JGR, 115, doi: 10.1029/2009 JA 14775. (2010)